\documentclass[3p,times]{elsarticle}

\usepackage{amssymb}
\usepackage{amsmath}

\usepackage[figuresright]{rotating}

\usepackage{subfig}

\usepackage{graphicx}
\usepackage{dcolumn}
\usepackage{bm}
\usepackage{epstopdf}
\usepackage{epsfig}
\usepackage[colorlinks = true,
linkcolor = blue,
urlcolor  = blue,
citecolor = blue,
anchorcolor = blue]{hyperref}
\usepackage{url}
\usepackage[mathlines]{lineno}
\linenumbers\relax 

\usepackage{amssymb}
\usepackage{amsthm}
\usepackage{ulem}

\usepackage[figuresright]{rotating}
\usepackage{subcaption}

\usepackage{graphicx}
\usepackage{dcolumn}
\usepackage{bm}
\usepackage{epstopdf}
\usepackage{epsfig}
\usepackage[colorlinks = true,
linkcolor = blue,
urlcolor  = blue,
citecolor = blue,
anchorcolor = blue]{hyperref}
\usepackage{url}
\usepackage[mathlines]{lineno}
\linenumbers\relax 

\theoremstyle{remark}

\usepackage{xcolor}

\AtBeginDocument{\colorlet{default}{.}}

\begin{document}
	\nolinenumbers

\begin{frontmatter}




\title{New logarithmic power nonlinear Schr\"odinger equations with super-Gaussons}


    	\author[au1]{Hadi Susanto} \ead{hadi.susanto@yandex.com} \address[au1]{Department of Mathematics, Khalifa University, PO Box 127788, Abu Dhabi, United Arab Emirates}

\begin{abstract}
We introduce a new class of nonlinear Schr\"odinger equations with a logarithmic--power nonlinearity that admits exact localized solutions of super-Gaussian form. The resulting stationary states possess flat-top profiles with sharp edges and are referred to as {super-Gaussons}, in analogy with the Gaussian Gaussons of the classical logarithmic NLS (log-NLS).
{The model, which we call the logarithmic-power NLS (logp-NLS), is parameterized by an exponent $p \geq 1$ that controls the degree of flatness of the soliton core and the sharpness of its decay. Mathematically, $p$ interpolates between the standard log-NLS ($p=1$) and increasingly flat-top profiles as $p$ increases, while physically it governs the stiffness of an underlying logarithmic--power compressibility law.}
The proposed equation is constructed so as to admit super-Gaussian stationary states and can be interpreted a posteriori within a generalized pressure-law framework, thereby extending the log-NLS. We investigate the dynamics of super-Gaussons in one spatial dimension through numerical simulations for various values of $p$, demonstrating how this parameter regulates both the internal structure of the soliton and its collision dynamics. The logp-NLS thus generalizes the standard log-NLS by admitting a broader family of localized states with distinctive structural and dynamical properties, suggesting its relevance for flat-top solitons in nonlinear optics, Bose--Einstein condensates, and related nonlinear media.
\end{abstract}


\end{frontmatter}



\section{Introduction}

The nonlinear Schr\"odinger (NLS) equation is one of the most fundamental models in nonlinear science, governing the propagation of waves in dispersive media. Its applications span optics \cite{hasegawa2003optical,kivshar2003optical}, Bose–Einstein condensates (BECs) \cite{kevrekidis2008emergent}, hydrodynamics \cite{pelinovsky2008extreme}, and plasma physics \cite{berge1998wave}. As a universal envelope equation for weakly nonlinear dispersive systems, it captures the interplay between nonlinearity and dispersion that leads to the formation of solitons and other coherent structures \cite{scott2003nonlinear}. The canonical cubic nonlinearity ($\sim |\psi|^{2}\psi$) has been studied extensively due to its integrability in one dimension \cite{shabat1972exact}, its tractability, and its relevance to Kerr-type optical media. However, various modifications of the cubic NLS have been proposed to describe physical regimes where the cubic model is insufficient \cite{salasnich2002effective}.

Among such modifications, the logarithmic nonlinearity ($\sim \ln|\psi|^{2}\,\psi$) occupies a distinguished place. First introduced by Rosen \cite{rosen1969dilatation} and independently by Bia{\l}ynicki-Birula and Mycielski \cite{bialynicki1975wave,bialynicki1976nonlinear}, the logarithmic NLS (log-NLS) arose as a natural extension of the standard NLS under a set of physical and mathematical constraints. These include: (i) separability of non-interacting subsystems, {, meaning that if a composite system consists of independent subsystems with no mutual interaction and the initial state is factorized, then the exact solution of the log-NLS remains factorized as a product of the corresponding single-subsystem solutions}; (ii) consistency with Planck’s relation for stationary states, requiring a linear energy–frequency relation; and (iii) stability in the sense of Poincar\'e, guaranteeing vanishing stress tensors for stationary solutions. {Unlike generic power-law nonlinearities, the exact multiplicative separability is exceptional and fails for most nonlinear Schr\"odinger models} \cite{bialynicki1975wave,bialynicki1976nonlinear}. The resulting log-NLS admits exact localized solutions called {Gaussons}, characterized by Gaussian profiles. These solutions interact inelastically \cite{oficjalski1978collisions,jakubowski1996can}, signaling the non-integrable nature of the equation. Properties of Gaussons and higher-dimensional excited states, including the influence of external potentials, were studied extensively in \cite{bialynicki1979gaussons}.

Although originally proposed as a conservative quantum model \cite{bialynicki1975wave}, the physical validity of the log-NLS was challenged after negative results from neutron interferometry experiments \cite{shull1980search,gahler1981neutron}. Alternative interpretations followed: Hefter \cite{hefter1985application} argued that the log-NLS may apply to extended objects rather than point-like particles, while Brasher \cite{brasher1991nonlinear} connected the logarithmic term with information-theoretic and thermodynamic arguments, relating it to temperature-dependent effects in many-particle systems. De Martino et al.\ \cite{de2003logarithmic} derived the log-NLS from hydrodynamic conservation laws under an assumption about the pressure tensor, providing a fluid-mechanical interpretation. In another direction, a stationary log-NLS has been shown to approximate Newton’s cradle dynamics, where Hertzian bead interactions yield effective logarithmic forces \cite{james2013breather,chatterjee1999asymptotic}. More recently, logarithmic nonlinearities have been linked to cavitation instabilities in liquid helium near critical negative pressures \cite{scott2019resolving,zloshchastiev2022resolving}, reinforcing the idea that such nonlinearities represent nontrivial physical compressibility laws.

Beyond physical modeling, the log-NLS has attracted considerable mathematical attention. The singular structure of the logarithm requires specialized regularization schemes for numerical computation \cite{bao2019error,bao2019regularized}, and the equation has been a testbed for developing new analytical methods \cite{carles2022logarithmic}. Its ground states are orbitally stable \cite{cazenave1983stable,ardila2016orbital}, even under harmonic confinement \cite{ardila2019logarithmic}, and variants with additional nonlinearities have been proposed \cite{wazwaz2020bright,liu2021two,al2024novel}. { Carles et al.\ \cite{carles2025ground} showed that the Gausson is obtained as the limiting ground state of the nonlinear Schr\"odinger equation with power-type nonlinearity in the zero-power limit.}  These studies highlight that the log-NLS sits at the crossroads of physics and mathematics: it is both a phenomenological model for unusual nonlinear responses and a mathematically rich equation with tractable localized solutions.

However, a limitation of the log-NLS is that its localized solutions are strictly Gaussian. In many physical contexts, the self-trapped states are not Gaussian but instead exhibit {flat-top} or super-Gaussian profiles \cite{kusdiantara2024analysis,otajonov2020variational,katsimiga2023interactions,barshilia2024quantum,baizakov2009solitons,adriano2025exponential,zeng2019purely}. These states feature nearly constant amplitude over a finite core and sharp decay at the edges, a structure that cannot be captured by either the cubic NLS or the pure log-NLS. To address this, we introduce a new class of logarithmic–power NLS equations (logp-NLS), defined by
\begin{equation}
	i \frac{\partial \psi}{\partial t} = - \Delta \psi - 4p^{2}\left(\ln^{2}{|\psi|}\right)^{\tfrac{p-1}{2p}}
	\left[\ln{|\psi|} + \Big(1 + \tfrac{d-2}{2p}\Big)\right]\psi,
	\label{eq:logpnls}
\end{equation}
{where \(\mathbf{x}=(x_1,x_2,\dots,x_d)\in\mathbb{R}^d\) is the spatial coordinate in $d$ dimensions and $\Delta = \sum_{i=1}^{d} \frac{\partial^2}{\partial x_i^2}$}. The logp-NLS admits as exact stationary solutions the super-Gaussian profiles
\begin{equation}
	\psi(\mathbf{x},t) = e^{-r^{2p}}\,e^{i\theta}, \qquad r = \sqrt{\sum_{i=1}^d x_i^2},\; p\geq1,\; \theta\in\mathbb{R}.
	\label{eq:supergauss}
\end{equation}
We refer to these solutions as {super-Gaussons}, in analogy with the Gaussian Gaussons of the log-NLS. {Here and in what follows, the term ``stationary solutions'' refers to standing localized states that are spatially immobile in the laboratory frame. Owing to the Galilean invariance of the logp-NLS in arbitrary spatial dimension $d$, traveling super-Gaussons are obtained by a standard Galilean boost of these stationary profiles. Consequently, the velocity does not enter the stationary problem, and no intrinsic speed–amplitude relation exists beyond that imposed by this symmetry. Particularly, if $\psi(\mathbf{x},t)$ is a solution, then for any constant velocity $\mathbf v\in\mathbb{R}^d$ the Galilean boost
\[
\psi_{\mathbf v}(\mathbf{x},t)
=
\psi(\mathbf{x}-\mathbf v t,t)
\exp\!\left(
i\left[\tfrac12\,\mathbf v\cdot\mathbf x-\tfrac14|\mathbf v|^2 t\right]
\right)
\]
is also a solution.}

The form of the nonlinearity in \eqref{eq:logpnls} is not arbitrary: it is determined by requiring that the radial Laplacian balance in dimension \(d\) admits \eqref{eq:supergauss} as an exact stationary state. Equivalently, the equation can be derived from a variational (Hamiltonian) principle with a generalized pressure law whose enthalpy involves both logarithmic and power-type dependence on the density. {From a physical standpoint, the logp-NLS may be viewed as a generalized hydrodynamic model of dispersive media in which the nonlinear response is described by an effective pressure (or compressibility) law, leading naturally to flat-top stationary states. This framework provides a continuum description of super-Gaussian localization that extends the classical log-NLS while preserving a Hamiltonian structure.}

{ 
A brief comment is in order concerning the role of the exponent $p$ in Eqs.~(\ref{eq:logpnls})-(\ref{eq:supergauss}). Varying $p$ affects several observable features of the stationary super-Gausson profiles, including the flatness of the core, the effective width, and the sharpness of the decay. These features, however, are not independently tunable: the parameter $p$ encodes a single modeling choice associated with the stiffness of the underlying logarithmic–power nonlinearity, and changes in the profile occur in a correlated manner. In particular, $p$ should not be viewed as an external control parameter for wave propagation, but rather as defining a family of related models interpolating between the Gaussian Gausson case ($p=1$) and increasingly flat-top states for $p>1$. A detailed hydrodynamic interpretation of how this stiffness enforces flat-top localization is provided in Sec.~\ref{sec3}.
}

In this work, we study the logp-NLS equation \eqref{eq:logpnls}, focusing on its stationary solutions and their dynamics. We show that the exponent \(p\) controls the degree of flatness of the soliton core, with the classical log-NLS corresponding to the Gaussian case \(p=1\). We introduce the family of super-Gaussons \eqref{eq:supergauss}, explore their stability properties, and analyze their dynamical behavior through direct numerical simulations in one dimension, {even though the stationary super-Gaussons are derived in arbitrary dimension under radial symmetry}.

\section{New NLS models} 

In radial (polar) coordinates, the Laplacian in the logp-NLS \eqref{eq:logpnls} takes the form
\begin{equation}
	\Delta = \frac{\partial^2}{\partial r^2} + \frac{d-1}{r}\frac{\partial}{\partial r} + \frac{1}{r^2}\nabla^2_{\Omega},
	\label{lap2}
\end{equation}
where \(\nabla^2_{\Omega}\) denotes the angular part of the Laplacian.  {Although the logp-NLS is formulated in arbitrary spatial dimension $d$, the use of polar (radial) coordinates in Eq.~(\ref{lap2}) is introduced solely to derive radially symmetric stationary solutions of super-Gaussian form. No restriction to a particular spatial dimension is imposed at this stage. In what follows, $d$ is kept general when discussing the structure of the equation and its stationary states.}

We seek localized stationary states of the form of a super-Gaussian function \eqref{eq:supergauss}. Applying the Laplacian operator \eqref{lap2} to the Ansatz \eqref{eq:supergauss} yields
\begin{equation}
	\Delta \psi = -4p^{2}\, r^{2p-2}\left[-\,r^{2p} + \Big(1+\tfrac{d-2}{2p}\Big)\right]\psi.
\end{equation}
Introducing the logarithmic relations
\begin{equation}
	-\,r^{2p} = \ln|\psi|, 
	\qquad 
	r^{2p-2} = \big(\ln^{2}|\psi|\big)^{\tfrac{p-1}{2p}},
	\label{eq:nonlinearity}
\end{equation}
we obtain the logp-NLS \eqref{eq:logpnls}. By construction, the super-Gaussian profile \eqref{eq:supergauss} is an exact stationary solution.  

The model \eqref{eq:logpnls} generalizes the log-NLS in the following sense. When \(p=1\), the exponent vanishes, and \eqref{eq:logpnls} reduces to the standard logarithmic NLS \cite{bialynicki1975wave,bialynicki1976nonlinear,bialynicki1979gaussons}.  
In the opposite limit \(p\gg 1\), the factor \((\ln^{2}|\psi|)^{(p-1)/(2p)}\) may be approximated at a formal level by \(|\ln|\psi||\), yielding
\begin{equation}
	i \frac{\partial \psi}{\partial t} = - \Delta \psi - 4p^{2}\,|\ln|\psi||\,
	\Big(\ln|\psi| + 1 + \tfrac{d-2}{2p}\Big)\psi.
	\label{eq:logpnls2}
\end{equation}
In this regime, the super-Gausson approaches a flat-top profile resembling a cylinder,
\begin{equation}
	\psi(r) \approx 
	\begin{cases}
		e^{i\theta}, & r<1, \\[6pt]
		0, & r>1,
	\end{cases}
	\label{sup2}
\end{equation}
with sharp edges at the boundary \(r=1\).

Following the general framework established in \cite{bialynicki1976nonlinear}, the logp-NLS inherits key structural properties of NLS-type equations: it admits a Lagrangian formulation, conserves the norm, and is invariant under Galilean transformations. However, the special properties of the pure log-NLS, i.e., separability of non-interacting subsystems, the exact Planck-type relation for stationary states, and Poincar\'e stability, are lost for \(p>1\). This reflects the fact that while the logp-NLS preserves the Hamiltonian structure of the log-NLS, it describes a broader class of self-trapped states that no longer share the exceptional symmetries of the Gaussian Gausson case.

\section{Variational structure and Hamiltonian formulation}\label{sec3}

The logarithmic–power nonlinear Schr\"odinger equation (logp-NLS) derived above can be placed in the standard Hamiltonian framework of NLS-type models. This formulation guarantees conservation laws and clarifies its interpretation in terms of generalized compressibility and pressure laws.


The logp-NLS \eqref{eq:logpnls} arises as the Euler–Lagrange equation of the action
\begin{equation}
S[\psi] \;=\;\int_{\mathbb{R}\times\mathbb{R}^{d}} 
\left\{ \frac{i}{2}\big(\psi\,\partial_t \psi^{*}-\psi^{*}\,\partial_t \psi\big) 
- |\nabla \psi|^{2} - G(|\psi|^{2}) \right\} dx\,dt,
\end{equation}
where the internal energy density \(G(\rho)\) in Madelung variables \(\psi=\sqrt{\rho}\,e^{iS}\) satisfies
\begin{align}
G'(\rho) = f(\rho)&=-4p^{2}\,\big(\ln^{2}\sqrt{\rho}\big)^{\tfrac{p-1}{2p}}
\left(\ln\sqrt{\rho} + 1 + \tfrac{d-2}{2p}\right)\nonumber\\
&= -\alpha_p\,|\ln\rho|^{\beta}\Big(\tfrac12\ln\rho + c_{d,p}\Big), \label{Gf}
\end{align}
where
\[\beta:=\frac{p-1}{p}\in(0,1),\;
\alpha_p:=4p^2\,2^{-\beta},\;
c_{d,p}:=1+\frac{d-2}{2p}.
\]
The Hamiltonian functional (energy) is
\begin{equation}
E[\psi] \;=\;\int_{\mathbb{R}^{d}} \left( |\nabla \psi|^{2} + G(|\psi|^{2}) \right)\,dx,
\end{equation}
which is conserved by Noether’s theorem. Additionally, \eqref{eq:logpnls} also conserves the mass (or particle number) $M[\psi]=\int_{\mathbb{R}^{d}}|\psi|^{2}\,dx$ and the momentum $\mathbf{P}[\psi]=\Im\int_{\mathbb{R}^{d}}\psi^{*}\nabla\psi\,dx$. It is also invariant under Galilean transformations.


In the Madelung variables, the governing equation \eqref{eq:logpnls} gives the compressible quantum fluid system \cite{abid2003gross}
\begin{align}
&\partial_t \rho + \nabla\cdot(\rho \nabla S) = 0, \quad 
\partial_t S + \tfrac12|\nabla S|^2 + h(\rho) - \frac{\Delta \sqrt{\rho}}{\sqrt{\rho}} = 0,
\end{align}
where the enthalpy satisfies
\begin{equation}
h(\rho)=f(\rho), 
\end{equation}
The associated pressure law and the barotropic sound speed are respectively given by
\begin{equation}
P(\rho)=\rho f(\rho)-G(\rho), \qquad c_s^2(\rho)=P'(\rho)=\rho f'(\rho).
\end{equation}
Thus, the logp-NLS corresponds to a barotropic fluid with a logarithmic–power compressibility law. Let $u=\ln\rho$. Accordingly, we have 
\begin{align}
f'(\rho)
&= -\frac{\alpha_p}{\rho}\left[
\beta\,|u|^{\beta-1}\operatorname{sgn}(u)\Big(\tfrac12 u + c_{d,p}\Big)
+\frac12\,|u|^{\beta}
\right].
\label{eq:prime}
\end{align}

\paragraph{Behavior near the flat-top level}
The stationary super-Gausson solutions are normalized so that $|\psi|=1$ corresponds to the plateau of the profile. For $\rho=|\psi|^{2}$ close to unity, let $\eta=\rho-1$, such that $u=\eta+O(\eta^{2})$. Then, using \eqref{eq:prime},
\begin{align}
f'(\rho)
&= -\frac{\alpha_p}{\rho}\left[
\beta\,c_{d,p}\,\operatorname{sgn}(\eta)\,|\eta|^{\beta-1}
+\frac12\,|\eta|^{\beta}
+O\!\big(|\eta|^{\beta}\big)
\right].
\label{eq:fprime-near1}
\end{align}
Since $\beta<1$, the exponent $\beta-1$ is negative, and therefore 
$|f'(\rho)|$ and $|P'(\rho)|=\rho|f'(\rho)|$ {diverge} as $\rho\to 1^{\pm}$, 
with opposite signs on the two sides of the plateau. 
In contrast, $f(\rho)\to 0$ as $\rho\to1$, so the nonlinear term in the stationary equation 
$\Delta\phi=f(\rho)\phi$ nearly vanishes in the core region. {
This mathematical structure provides a precise explanation for the formation of a flat-top soliton. While the nonlinear contribution $f(\rho)$ itself vanishes as $\rho \to 1$, its derivative $f'(\rho)$ diverges, implying that the pressure derivative $P'(\rho)=\rho f'(\rho)$ becomes unbounded in the vicinity of the plateau. As a consequence, arbitrarily small deviations of the density from $\rho=1$ generate a disproportionately large restoring response. In hydrodynamic terms, the effective compressibility $\kappa(\rho)\propto 1/P'(\rho)$ tends to zero as $\rho\to1^\pm$, so the medium behaves as locally incompressible in the bulk of the soliton.} 

{This stiffness enforces a nearly constant density in the core and localizes the transition to the edges within a narrow region of strong restoring forces. While the resulting profiles decay superexponentially due to their super-Gaussian form, the sharp boundaries of the super-Gausson arise from the rapid increase of $|P'(\rho)|$ away from the plateau together with the regularizing effect of the quantum pressure term.
}

\paragraph{Behavior away from the plateau}
For \(\rho\to0\) and \(\rho\to\infty\), 
we have
\begin{equation}
c_s^2(\rho)=P'(\rho)
= -\alpha_{p}\!\left[\tfrac{1}{2}(1+\beta)\,|\ln\rho|^{\beta}+o(|\ln\rho|^{\beta})\right],
\qquad |\ln\rho|\to\infty.\label{eq:cs2}
\end{equation}
Hence, the magnitude of the pressure derivative grows as $|P'(\rho)|\sim|\ln\rho|^{\beta}$ in both low- and high-density regimes. 
This increase in $|P'|$ indicates that the restoring force strengthens rapidly as the density departs from the equilibrium level, 
enforcing confinement and producing the sharp edges of the super-Gausson. 
Physically, the medium becomes progressively less compressible away from the flat-top core, 
so that the combined effect of the nonlinear pressure and quantum pressure terms sustains a self-trapped state 
with steep boundaries and a nearly uniform interior.

\paragraph{Sign and focusing character}
Equations \eqref{eq:prime} show that \(c_s^2(\rho)=P'(\rho)\) is negative for sufficiently large \(|u|=|\ln\rho|\) on both sides and changes sign across \(\rho=1\) due to the \(\operatorname{sgn}(u)\) term. This reflects the {focusing} character of the nonlinearity in \eqref{Gf}. In the hydrodynamic picture, negative \(P'(\rho)\) corresponds to an imaginary acoustic speed (modulational tendency), while the quantum pressure term balances it to sustain a localized equilibrium (the super-Gausson). The singular behavior of \(c_s^2(\rho)\) from \eqref{eq:fprime-near1} leads to the flat-top plateau (weak restoring force at \(\rho\approx1\)) and the sharp edges (strong response for \(|\ln\rho|\gg1\)).


From this viewpoint, the logp-NLS is not only an ad hoc modification of the log-NLS, but also a Hamiltonian fluid model with an enthalpy depending on density through a logarithmic power law. The exponent \(p\) controls the degree of flattening of the soliton core, and the logp-NLS provides a systematic Hamiltonian extension of the log-NLS that captures a broader class of physically relevant self-trapped states while preserving the variational and hydrodynamic structure.

\section{Numerical simulations} 

{In this section, we restrict our attention to the one-dimensional case ($d=1$), which allows a tractable investigation of the super-Gausson solutions \eqref{eq:supergauss}. All numerical simulations reported below focus on their stability and nonlinear dynamics.}

\subsection{Linear stability analysis}

\begin{figure}[tbhp]
	\centering
	\subfloat[]{\includegraphics[width=0.48\textwidth]{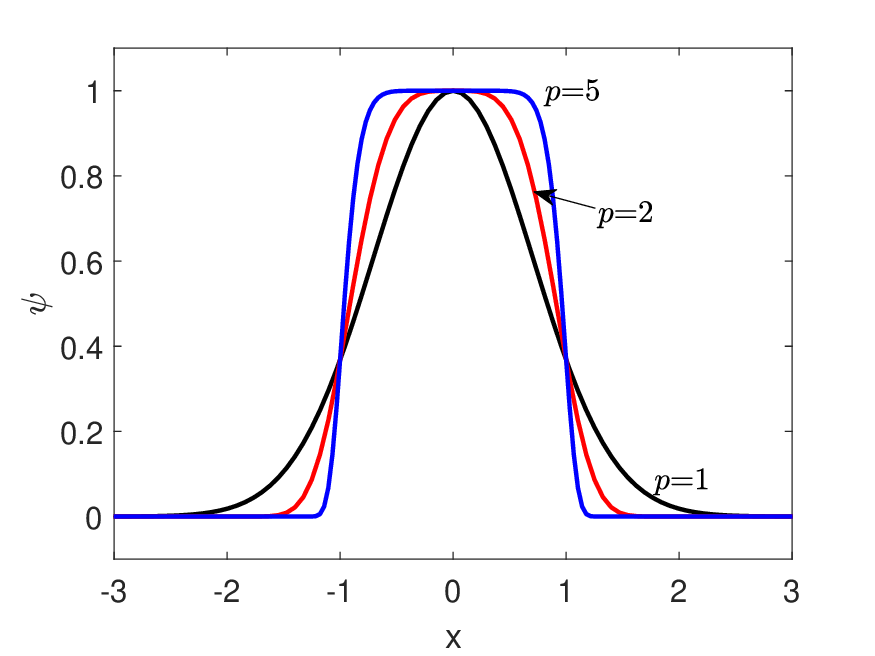}}
	\subfloat[]{\includegraphics[width=0.48\textwidth]{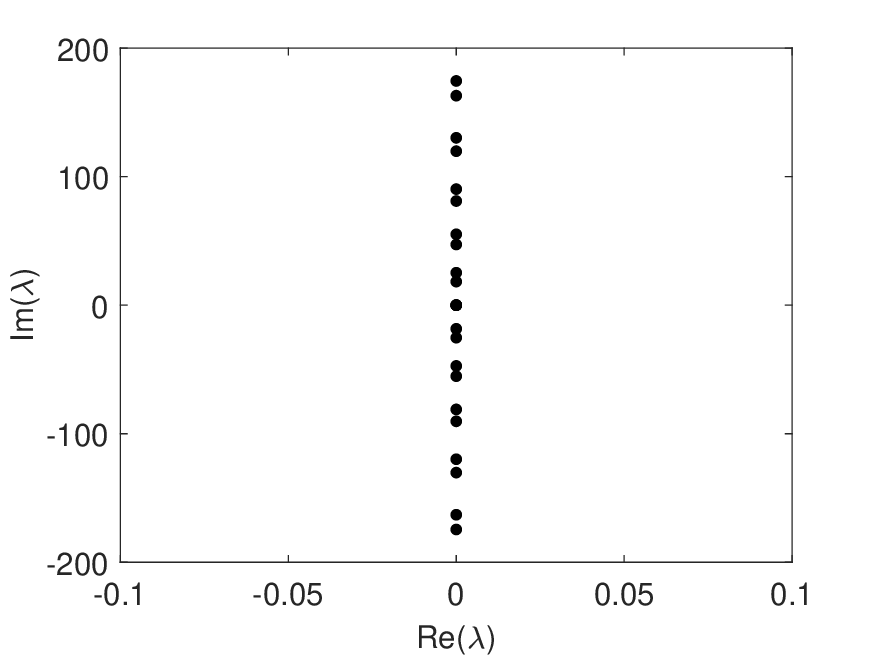}}
	\caption{(a) Stationary super-Gausson profiles $\psi$ for several values of the exponent $p$, illustrating the transition from the Gaussian profile ($p=1$) to increasingly flat-top shapes as $p$ increases. (b) Spectrum of the linearized operator around the super-Gausson for $p=2$, shown in the complex plane. The spectrum consists solely of discrete eigenvalues, with no continuous spectrum present. All eigenvalues lie on the imaginary axis, indicating linear stability of the super-Gausson.}
	\label{fig:profiles}
\end{figure}

{Consider the stationary super-Gausson
\[
\psi_0(x,t)=\mathrm{e}^{-(x^2)^p},
\]
and introduce a small perturbation in the form
\[
\psi(x,t)=\psi_0(x)+\varepsilon\big[v(x,t)+i w(x,t)\big], \qquad 0<\varepsilon\ll1.
\]
Substituting this ansatz into the logp-NLS equation \eqref{eq:logpnls} and expanding to first order in \(\varepsilon\), we obtain a linear system for the perturbation components \(v\) and \(w\),
\[
\partial_t
\begin{pmatrix}
v\\
w
\end{pmatrix}
=
\underbrace{
\begin{pmatrix}
0 & -\partial_x^2 + V_w(x)\\
\partial_x^2 - V_v(x) & 0
\end{pmatrix}}_{\mathcal{L}}
\begin{pmatrix}
v\\
w
\end{pmatrix}.
\]
The effective potentials \(V_w\) and \(V_v\) arise from the linearization of the nonlinear term $f(|\psi|^2)\psi$ about \(\psi_0\). Writing \(f'(\rho)\) for the derivative of the nonlinear coefficient with respect to \(\rho=|\psi|^2\), one finds
\[
V_w(x)=f(\psi_0^2),
\qquad
V_v(x)=f(\psi_0^2)+2\psi_0^2 f'(\psi_0^2),
\]
which is the standard structure for Hamiltonian NLS-type equations. Substituting the explicit expression for \(f\) associated with the logp-NLS and using \(\psi_0(x)=\mathrm{e}^{-(x^2)^p}\) yields the potentials }
\begin{align}
V_w(x) &= 4p^2 (x^2)^{p-1}\left(\tfrac{1}{2p}-1+(x^2)^p\right), \\
V_v(x) &= (x^2)^{p-1}\left( 4p^2(x^2)^{p} - 12p^2 + 6p + (4p^2-6p+2)(x^2)^{-p}\right).
\end{align}
The solution is said to be linearly stable if the spectrum of the operator \(\mathcal{L}\) lies on the imaginary axis.  

When \(p=1\), one recovers \(V_w=4x^2-2\) and \(V_v=4x^2-6\), so that \(\mathcal{L}\) reduces to an operator closely related to the quantum harmonic oscillator \cite{bender1999advanced,ciftci2003asymptotic}. For \(p>1\), the potentials become higher-degree polynomials, so that \(\mathcal{L}\) describes an anharmonic oscillator; see \cite{turbiner2023quantum} for a review. Remarkably, for \(p>1\), the potential \(V_v\) also contains a singular term \(\sim 1/x^2\). While this potential diverges as \(x\to 0\), such singularities are not pathological in quantum models, similar to the Coulomb potential or the Dirac delta well \cite{coon2002anomalies,essin2006quantum,cisneros2007comment,harrell1977singular,ciftci2003asymptotic}. 

{Figure~\ref{fig:profiles} illustrates both the structure and the linear stability of the super-Gausson solutions. Panel~(a) shows representative stationary profiles $|\psi(x)|$ for several values of the exponent $p$, highlighting the progressive flattening of the soliton core and the sharpening of the edges as $p$ increases. Panel~(b) displays the spectrum of the linearized operator $\mathcal{L}$ about the super-Gausson for $p=2$, shown in the complex plane. The spectrum is obtained by discretizing the spatial domain $-L/2 \leq x < L/2$ with $L=13\pi$ and $N=2^{10}$ grid points (i.e., the spatial discretization is $\delta x\approx0.0399$). To ensure numerical robustness, a mild smoothing of the logarithmic–power nonlinearity is applied, as described in the following subsection (see Eq.~\eqref{ln_delta}). The resulting spectrum consists exclusively of discrete eigenvalues, with no continuous spectrum observed. Moreover, all eigenvalues lie on the imaginary axis, so that no modes with a positive real part are present, providing numerical evidence for the linear stability of the super-Gausson.}

\subsection{Numerical methods for the nonlinear dynamics}

{To investigate the nonlinear dynamics of the logp-NLS \eqref{eq:logpnls}, we perform direct numerical simulations in one spatial dimension. Spatial derivatives are approximated spectrally using a Fourier pseudospectral method, implemented via the fast Fourier transform (FFT), with periodic boundary conditions applied to the computational domain. Time integration is carried out using a classical fourth-order Runge-Kutta (RK4) scheme with fixed time step $\delta t=\delta x/100$. This combination provides high-order accuracy for smooth solutions and is widely used for nonlinear Schr\"odinger-type equations. Additional numerical details and regularization procedures are described below.}

Unlike polynomial or pure logarithmic nonlinearities, the logarithmic--power term introduces a singular derivative at $|\psi|=1$. From the analytical expansion near $\rho = |\psi|^{2} = 1$, see \eqref{eq:fprime-near1}, we have $|f'(\rho)| \to \infty$ as $\rho \to 1$. As a consequence, even very small perturbations—such as round-off errors or high-frequency components in $|\psi|$—can induce disproportionately large variations in the nonlinear phase term. This reflects an intrinsic stiffness of the underlying partial differential equation, rather than a deficiency of a particular numerical discretization. {In practice, this singular stiffness severely constrains time integration: explicit schemes exhibit strong step-size restrictions and may suffer numerical blow-up when the solution approaches the flat-top level $|\psi|\approx1$, while implicit schemes do not fundamentally remove the difficulty, as the nonlinear Jacobian becomes ill-conditioned near $\rho=1$.} Expanding the nonlinearity \eqref{eq:nonlinearity} around $p=1$ further highlights this singular structure,
\[
\big(\ln^2|\psi|\big)^{\tfrac{p-1}{2p}}
= 1 + (p-1)\ln\!\big(|\ln|\psi||\big)+\dots,
\]
which diverges logarithmically as $|\psi| \to 1$. Hence, the observed numerical instabilities are not purely algorithmic, but are a direct manifestation of the singular stiffness inherent in the analytical model itself.

To control this stiffness, we adopt the regularization strategy of \cite{bao2019regularized} and extend it to the logp-NLS. We replace the logarithm with 
\begin{equation}
\ln_\delta(|\psi|):=\tfrac{1}{2}\ln(|\psi|^{2}+\delta), \qquad \delta>0, \label{ln_delta}
\end{equation}
and define the regularized nonlinear term
\begin{equation}
f_\delta(\rho) = -4p^{2}\,\Big(\ln_{\delta}^{2}|\psi|\Big)^{\tfrac{p-1}{2p}}
\Big(\ln_{\delta}|\psi| + 1 + \tfrac{d-2}{2p}\Big).
\end{equation}
The corresponding equation
\[
i\,\partial_t\psi = -\Delta\psi + f_\delta(\rho)\psi
\]
approximately preserves the Hamiltonian structure, sufficient to maintain the qualitative soliton dynamics 
and ensures that $f_\delta'(\rho)$ remains bounded for all $\rho>0$. 
The typical value $\delta\sim10^{-15}$ suffices to suppress spurious oscillations while leaving the macroscopic soliton dynamics unchanged.

In addition to the analytic regularization above, we initialize with super-Gausson profiles of amplitude slightly below unity, $\max|\psi|<1$, to avoid the most singular region. With this modification, the simulations become stable over long propagation times and accurately reproduce soliton interactions and breathing dynamics. 

To this end, we first compute stationary localized states of the form
\[
\psi(x,t)=\Psi(x)e^{i\omega t},
\]
which satisfy
\[
\Psi'' -f(|\Psi|^2) 
\Psi + \omega\Psi 
= 0.
\]
As closed-form solutions are unavailable, we solve this boundary-value problem numerically by a Newton–Raphson iteration combined with spectral differentiation. The computed profiles are then used as initial conditions for time evolution under the regularized dynamics.

\subsection{Dynamics of collisions}

\begin{figure}[tbhp]
	\centering
	\subfloat[$p=1,\,v=0.2$]{\includegraphics[width=0.48\textwidth]{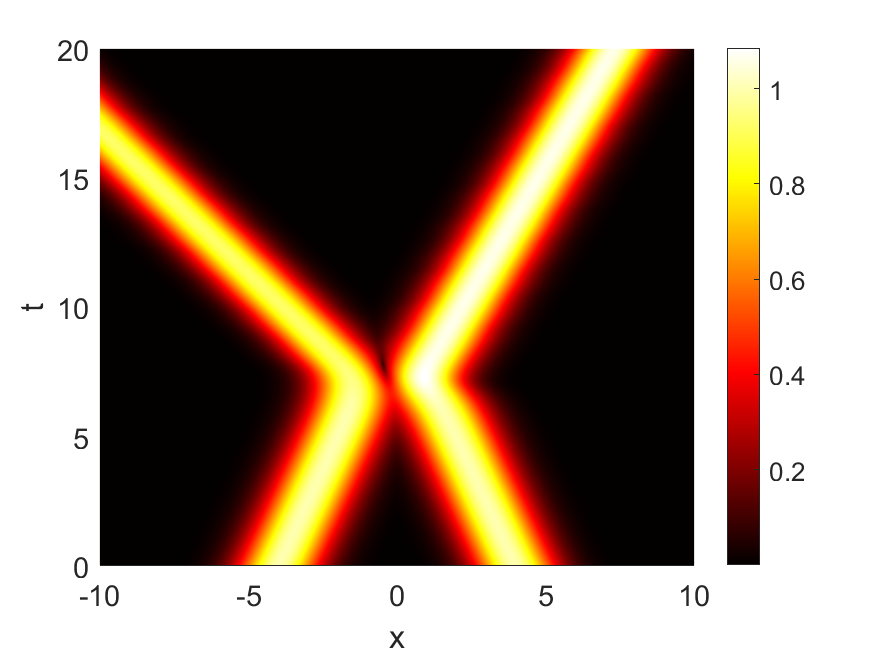}}
	\subfloat[$p=1,\,v=2.0$]{\includegraphics[width=0.48\textwidth]{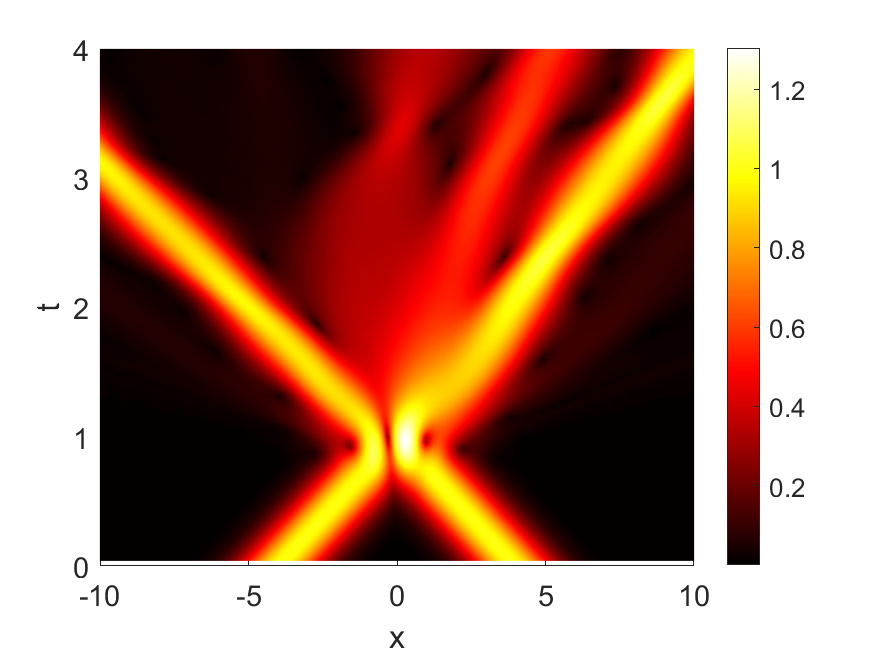}}\\
	\subfloat[$p=2,\,v=0.2$]{\includegraphics[width=0.48\textwidth]{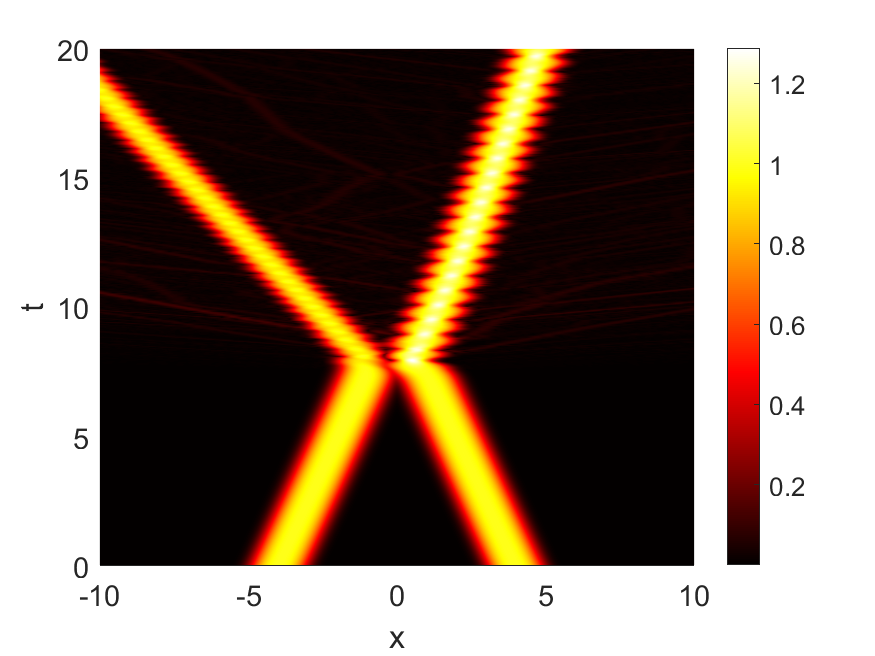}}
	\subfloat[$p=2,\,v=2.0$]{\includegraphics[width=0.48\textwidth]{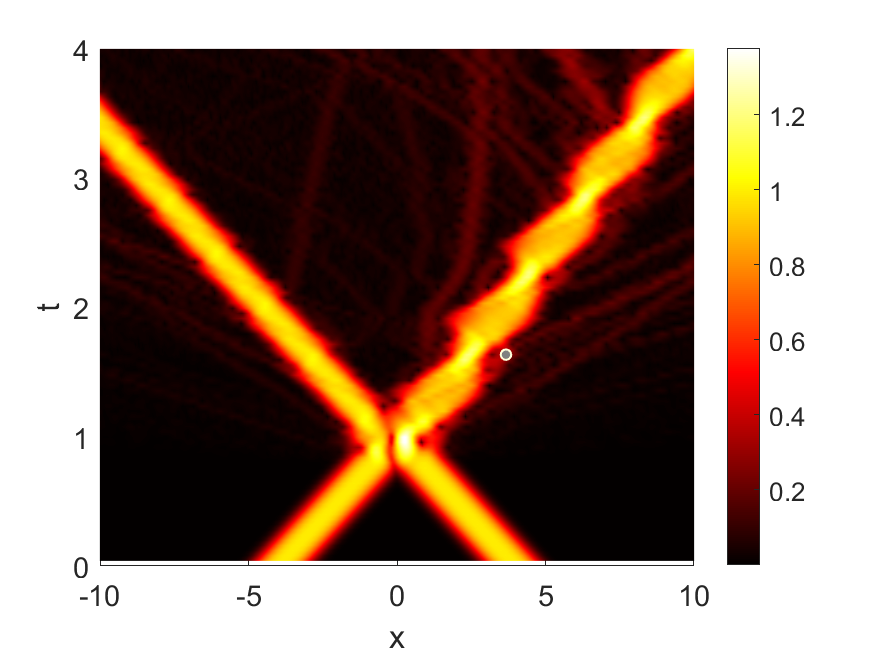}}\\
	\caption{Inelastic collisions of two super-Gaussons, initially separated, with $\omega=0.01$. 
	Shown is the top view of $|\psi(x,t)|^{1/2}$.}
	\label{fig1}
\end{figure}

In Fig.~\ref{fig1}, we present the temporal evolution of two colliding super-Gaussons for different values of the exponent \(p\). 
The initial condition is taken as 
\[
\psi(x,0) = \Psi(x+x_{0})\,e^{ivx} + \Psi(x-x_{0})\,e^{-i(vx-\pi/2)}, 
\qquad x_{0}\gg 1,
\]
so that the two solitons are well separated initially and possess a relative phase difference of \(\pi/2\).  The figure shows the evolution of the amplitude $|\psi(x,t)|^{2}$ during head-on collisions for several values of \(p\) and velocity \(v\). 
Panels~(a) and~(b) correspond to the classical log-NLS case ($p=1$), while panels~(c) and~(d) display the corresponding results for the logp-NLS with $p=2$. 
All simulations are performed using the regularized equation with $\omega=0.01$.  

At low collision velocities [Figs.~\ref{fig1}(a) and \ref{fig1}(c)], the solitons approach each other, overlap, and then separate with almost complete recovery of their initial profiles. 
Nevertheless, the outgoing waves exhibit persistent oscillations in width (breathing) and weak radiation, particularly for $p=2$, indicating that the collisions are {inelastic}. 
This behavior confirms that neither the log-NLS nor the logp-NLS is integrable, in contrast to the cubic NLS, where soliton collisions are perfectly elastic. 
The degree of inelasticity and the amplitude of the residual breathing increase with the nonlinearity exponent~$p$, as seen by comparing the $p=1$ and $p=2$ results.

For higher collision velocities [Figs.~\ref{fig1}(b) and \ref{fig1}(d)], stronger inelastic effects are observed in the $p=1$ case: the overlap region develops a series of oscillatory localized patterns and radiative fronts, indicating that part of the kinetic energy is transferred into internal breathing modes. 
The post-collision solitons remain localized but emerge with oscillating amplitude and width. 
Interestingly, these effects become less pronounced for larger~$p$, which is consistent with the stronger stiffness of the logarithmic–power nonlinearity near $|\psi|=1$. 
The enhanced rigidity of the flat-top solitons limits deformation and suppresses energy exchange during rapid impacts, resulting in comparatively more elastic collisions at higher velocities.

The velocity dependence of inelasticity may be understood from the hydrodynamic viewpoint as follows. At low velocities, the interaction time between solitons is long, allowing their flat-top cores to overlap significantly. 
The strong nonlinear pressure gradients near $\rho\simeq1$ then excite internal oscillations, leading to a pronounced exchange of energy between translational and internal modes, an effect that becomes stronger with increasing $p$. 
At high velocities, by contrast, the interaction time is short, and the same stiffness acts as an effective rigidity: the solitons behave like nearly incompressible droplets that pass through each other with limited deformation. 
Hence, the inelasticity is more pronounced for larger $p$ when the collision is slow, whereas for fast collisions, the trend reverses, and larger $p$ yields more elastic scattering.

Overall, Fig.~\ref{fig1} demonstrates that the inclusion of the logarithmic–power nonlinearity in the logp-NLS not only preserves the existence of localized solitary waves but also introduces a richer dynamical behavior. 
The interplay between the exponent $p$ and the collision speed $v$ controls the balance between deformation, internal excitation, and radiation, reflecting the generalized compressibility of the underlying medium.

\section{Conclusion} 

In this work, we introduced a new class of logarithmic–power nonlinear Schr\"odinger equations (logp-NLS) that admit exact soliton solutions in the form of super-Gaussian functions, which we referred to as super-Gaussons. We established the variational and hydrodynamic structure of the model and derived the corresponding eigenvalue problem governing linear stability, which reduces to a pair of coupled Schr\"odinger-type equations with anharmonic and singular potentials. Numerical computations provide evidence of spectral stability for the super-Gaussons considered here and revealed that collisions between them are inelastic, demonstrating that the logp-NLS is non-integrable. At the computational level, we identified numerical instabilities associated with the logarithmic–power nonlinearity, particularly near the singular point \(|\psi|=1\), and demonstrated that these can be partially mitigated by suitable regularizations and amplitude adjustments. 

Several open problems remain. A key direction is the analytical characterization of the spectrum of the linearized operator, which would provide a rigorous understanding of the stability and possible internal modes of super-Gaussons. A systematic analysis of their dynamical properties, including long-time behavior and multi-soliton interactions, is also of considerable interest. On the computational side, the development of robust and structure-preserving numerical schemes for logarithmic–power nonlinearities is essential. Most importantly, potential physical applications of the logp-NLS should be investigated: its generalized pressure law may suggest relevance to flat-top states in nonlinear optics, Bose–Einstein condensates, and fluid or granular media. Bridging the mathematical theory developed here with experimental settings would significantly advance the understanding of super-Gaussian solitons in real-world systems.

\section*{Data availability}
No data was used for the research described in the article.


\section*{Declaration of generative AI and AI-assisted technologies in the writing process}

During the preparation of this work, the author utilized Grammarly and ChatGPT to enhance language and readability. After using these tools/services, the author reviewed and edited the content as needed and takes full responsibility for the content of the publication.

\section*{Acknowledgements}

HS acknowledges support from Khalifa University through	a Competitive Internal Research Awards Grant (No.\ 8474000413/CIRA-2021-065) and Research \& Innovation Grants (No.\ 8474000617/RIG-S-2023-031 and No.\ 8474000789/RIG-S-2024-070).

\bibliographystyle{elsarticle-num}
\bibliography{references}







\end{document}